\documentclass[sigconf,balance=false]{acmart}
\usepackage{xfrac,xifthen,amsmath,amsfonts,url,mathrsfs,todonotes}
\usepackage{hyperref}
\usepackage{paralist}
\usepackage{xparse}
\newcommand{\FF}{\mathbb{F}}
\newcommand{\QQ}{\mathbb{Q}}
\newcommand{\RR}{\mathbb{R}}
\newcommand{\SK}{\mathscr{K}}
\newcommand{\SL}{\mathscr{L}}
\newcommand{\QQx}{\QQ[x]}
\newcommand{\og}{\overline{g}}
\newcommand{\op}{\overline{p}}
\newcommand{\ideal}[1]{(#1)}
\newcommand{\norm}[1][{\quo{L}{K}}]{N_{#1}}
\newcommand{\onto}{\twoheadrightarrow}
\newcommand{\quo}[2]{#1/#2}
\newcommand{\st}{\mid}
\DeclareMathOperator{\lc}{lc}
\DeclareMathOperator{\ord}{ord}
\DeclareMathOperator{\poly}{poly}
\DeclareMathOperator{\height}{height}
\newcommand{\term}[1]{\emph{#1}}
\DeclareMathOperator{\IntFact}{IntFact}
\DeclareMathOperator{\PolyFact}{PolyFact}
\DeclareMathOperator{\UGComp}{UnitGroupComp}
\DeclareMathOperator{\SFUnits}{SizeFundUnits}
\newtheorem*{EI}{Euler's identity}
\newtheorem{thm}{Theorem}
\newtheorem{obs}[thm]{Observation}
\newtheorem{defn}[thm]{Definition}
\newtheorem{prop}[thm]{Proposition}

\theoremstyle{remark}
\newtheorem{rem}[thm]{Remark}
\newenvironment{poc}[1][]{\begin{proof}[\ifthenelse{\equal{#1}{}}{Proof of correctness}{Proof of correctness of #1}]}{\end{proof}}

\NewDocumentEnvironment{comp_analysis}{o o}{\begin{proof}[\IfNoValueTF{#1}{Complexity analysis}{ \IfNoValueTF{#2}{Complexity analysis of #1}{Complexity analysis of #1 and #2}}]}{\end{proof}}

\usepackage{algorithm}
\usepackage[noend]{algorithmic}

\def\clap#1{\hbox to 0pt{\hss#1\hss}}

\def\mathrlap{\mathpalette\mathrlapinternal}

\def\mathrlapinternal#1#2{\rlap{$\mathsurround=0pt#1{#2}$}}

\begin{document}

%\title[Pourchet's theorem in action]{Pourchet's theorem in action: decomposing univariate nonnegative polynomials as sums of five squares}
\title{Pourchet's theorem in action: decomposing univariate nonnegative polynomials as sums of five squares}

\author{Victor Magron}
%\email{@}
%\email{}
\affiliation{%
  \institution{CNRS LAAS \& Institut de Mathématiques de Toulouse}
  \streetaddress{7 avenue du Colonel Roche}
  \city{Toulouse}
  \country{France}
  \postcode{F-31400}
}
\email{victor.magron@laas.fr}

\author{Przemys{\l}aw Koprowski}
\affiliation{%
  \institution{University of Silesia in Katowice, Institute of Mathematics}
  \streetaddress{ul. Bankowa 14}
  \city{Katowice}
  \country{Poland}
  \postcode{40-007}
}
\email{przemyslaw.koprowski@us.edu.pl}

\author{
  Tristan Vaccon}
\affiliation{Universit\'e de Limoges;
  \institution{CNRS, XLIM UMR 7252}
  \city{Limoges}
  \country{France}  
  \postcode{87060}  
}
\email{tristan.vaccon@unilim.fr}

\begin{abstract}
Pourchet proved in 1971 that every nonnegative univariate polynomial with rational coefficients is a sum of five or fewer squares. Nonetheless, there are no known algorithms for constructing such a decomposition. The sole purpose of the present paper is to present a set of algorithms that decompose a given nonnegative polynomial into a sum of six (five under some unproven conjecture or when allowing weights) squares of polynomials. Moreover, we prove that the binary complexity can be expressed polynomially in terms of classical operations of computer algebra and algorithmic number theory.
\end{abstract}


\maketitle
%%%%%%%%%%%%%%%%%%%%%%%%%%%%%%%%%%%%%%%%%%%%%%%%%%%%
% P.K.: THE NEXT LINE CUTS DOWN THE VERTICAL LENGTH
\addtolength{\abovedisplayskip}{-2bp}
%%%%%%%%%%%%%%%%%%%%%%%%%%%%%%%%%%%%%%%%%%%%%%%%%%%%
\section{Introduction}
\label{sec:intro}

Let $\QQ[x]$ and $\RR[x]$ denote the sets of univariate polynomials with rational and real  coefficients, respectively. 
Given a nonnegative polynomial $f \in \QQ[x]$, we consider the problem of decomposing $f$ as a sum of squares (SOS) of polynomials also lying in $\QQ[x]$, possibly with rational positive weights. 

This problem is not only of theoretical interest in the realm of real algebraic geometry but is also practically meaningful, for instance, to compute SOS-Lyapunov certificates to ensure the stability of a control system \cite{rantzer2000convexity}, certify polynomial approximations of transcendental functions  evaluated within computer programs \cite{chevillard2011efficient,magron2017certified},  verify formally polynomial inequalities via proof assistants \cite{besson2007fast,Magron2015,hales2017formal}. 
Since such proofs assistants have limited computational abilities, a typical approach is to rely on external tools providing SOS certificates of moderate bit size, so that further verification is not too time-consuming.
Therefore, we are particularly motivated in designing algorithms that output SOS certificates of reasonable bit size, possibly with a bit complexity being polynomial in the input data. 

~

\paragraph{\bf Related works}
%
%References to the univariate SOS case (real and complex, cite Santos-Mourrain), multivariate SOS case, gradient SOS, other types of certificates (SONC/SAGE). 

It is well-known that every nonnegative univariate polynomial in $\RR[x]$ can be decomposed as a sum of two squares. 
Very early research efforts have been focused on obtaining rational decompositions with the least number of needed squares.
Landau proved in \cite{landau1906darstellung} that each nonnegative polynomial in $\QQ[x]$ can be decomposed as a sum of at most eight polynomial squares in $\QQ[x]$. 
This result has been improved by Pourchet in \cite{pourchet1971representation}, where he proved that only five or fewer squares are needed. 
The proof of Pourchet's theorem heavily relies on the local-global principle, in particular the Hasse-Minkowski theorem, and at first glance, it is not straightforward to extract a constructive algorithm to output the desired SOS decomposition.  
For a presentation in English of Pourchet's theorem, we refer the reader to \cite[Chapter~17]{Rajwade1993}.
Later, Schweighofer derived in \cite{schweighofer1999algorithmische} an algorithm to produce SOS decompositions of polynomials with coefficients lying in any subfield of~$\RR$. 
Here the number of output squares does not exceed the degree of the input polynomial.
This recursive algorithm performs real root isolation and quadratic approximations of positive polynomials at each of the recursion steps, thus has an exponential bit complexity \cite{magron2019algorithms}.
Another algorithm derived in \cite[\S~5.2]{chevillard2011efficient} relies on approximating
complex roots of perturbed positive polynomials. 
In contrast to Schweighofer's algorithm, the input must be rational, and three additional squares can appear in the resulting decompositions, but the bit complexity happens to be polynomial \cite{magron2019algorithms}. 

Alternatively, one can obtain an approximate rational SOS decomposition by checking the feasibility of a semidefinite program (SDP), namely by solving a convex problem involving linear equalities and linear matrix inequalities; see the seminal works by Parrilo \cite{parrilo2000structured} and Lasserre \cite{lasserre2001global}.
The modern development of floating-point SDP solvers ensures that this task can be efficiently done in practice, but a post-processing step is mandatory to obtain an exact decomposition, either based on rounding-projection schemes \cite{peyrl2008computing}, or perturbation-compensation techniques \cite{magron2021exact}.
One can also directly compute exact algebraic solutions to such SDP programs \cite{henrion2019spectra}, but the related schemes have more limited scalability.
Note that all such SDP-based frameworks can be generally applied to prove the existence of SOS decompositions in the multivariate case. 

Such schemes based on rounding-projection or perturbation-compensation techniques have been extended to design and analyze algorithms producing positivity certificates for trigonometric polynomials \cite{magron2022exact}, sums of nonnegative circuits \cite{magron2023sonc}, and sums of arithmetic-geometric-exponentials \cite{magron2019exact}. 
Recently, some other generalizations have been studied; they include the  case of univariate polynomials sharing common real roots with another univariate polynomial  \cite{krick2021univariate}, or the case of multivariate  polynomials whose gradient ideals are zero-dimensional and radical \cite{gradsos}. 
The two algorithms from \cite{magron2019algorithms} providing univariate rational SOS decompositions  have been implemented in the {\tt RealCertify} Maple library \cite{magron2018realcertify}.

~

\paragraph{\bf Contributions}
To the best of our knowledge, no constructive algorithm implements Pourchet's theorem to decompose a given nonnegative univariate polynomial from $\QQ[x]$ into a sum of five squares of polynomials in $\QQ[x]$.
We also investigate the bit complexity trade-off between  decompositions involving a fixed number of squares (namely five) and the existing schemes analyzed in \cite{magron2019algorithms} where the decompositions involve a number of squares depending on the degree of the input.
The central contribution of our work is to present and analyze an algorithm (Algorithm \ref{alg:sixsquares}) to decompose a nonnegative univariate polynomial $f \in \QQ[x]$ into a sum of {\bf six} squares of polynomials, also  with rational coefficients. 
When allowing positive rational weights, this algorithm can be used to obtain a weighted rational sum of {\bf five} squares with rational coefficients.
This main algorithm relies on other procedures related to the number of squares involved in the decomposition of the input.
\begin{itemize}
\item First, we focus in Section \ref{sec:twosquares} on the case where $f$ is a sum of two squares and design Algorithm \ref{alg:sos2} to implement the related decomposition. The main step of the algorithm consists of solving a norm equation involving the leading coefficient of~$f$; 
\item Then, we handle in Section \ref{sec:foursquares} the case when $f$ is a sum of four squares and design Algorithm \ref{alg:sos4} to implement the related decomposition. 
This algorithm is based on decompositions of positive rational numbers as sums of four squares of rational numbers, Euler's identity, and another norm equation;
\item Section \ref{sec:fivesquares} focuses on the reduction to the four square case. 
This reduction is performed in Algorithm \ref{alg:reduce_helper} by examining the 2-adic valuations of the constant term and leading coefficient of $f$, and using a perturbation argument, similar to the one used in \cite[\S~5.2]{chevillard2011efficient};
\item Last but not least, let $f$ be a nonnegative univariate rational polynomial of degree $d$ with coefficients of maximal bitsize $\tau$.
We prove that Algorithm \ref{alg:sixsquares}
computes a decomposition of $f$ as a sum of 6 squares.
%For some $\mathscr{H}$ polynomial in $d,\tau, \inf (f),$ and $\inf (f_*)$, 
The expected bit complexity of this computation can be expressed polynomially
in terms of: \begin{inparaenum}
\item integer factorization,
\item factorization in $\mathbb{Q}[x]$,
\item computation of the unit group of a number field,
\item the size of a system of fundamental units of a number field,
\end{inparaenum}
all applied on parameters of size polynomial in $d,\tau$, and the minimal values of $f$ and its reciprocal. We write \textit{expected} as there are some \textbf{Las Vegas} subalgorithms in Section \ref{sec:foursquares}.
%$\inf (f),$ and $\inf (f_*)$.
The output will also have bitsize polynomial in 
the size of a system of fundamental units of a number field
with the same parameters.
%
%For some $\mathscr{H}$ polynomial in $d,\tau, \inf (f),$ and $\inf (f_*)$ such that Algorithm \ref{alg:sixsquares} uses an amount of  boolean operations 
%polynomial in $\IntFact(\tau)$, $\PolyFact(d,\mathscr{H})$, $\UGComp(2d,\mathscr{H})$, and $\SFUnits(2d,\mathscr{H})$
%and returns polynomials of bitsize bounded by a polynomial
%in $\SFUnits(2d,\mathscr{H})$, where 
%$\IntFact,\PolyFact,$ and $\UGComp$ stand for the cost of integer factorization,
%polynomial factorization and unit group compution in
%a number field and $\SFUnits$ stands for an upper bound on the bitsize
%of a system of fundamental units in a number field.
%\todo[inline]{Tristan: Is it a satisfying presentation of the complexity?}
%\todo[inline]{Victor: Yes, it is great. Since we did not define $\inf (f)$ and $\inf (f_*)$, I proposed a reformulation with words.}
%\todo[inline]{Victor: Later on the degree of $f$ is $n$ but we used $d$ in the description of the algorithms, so I changed them.}
\end{itemize}

\paragraph{\bf Complexity of ordinary operations in computer algebra and algorithmic number theory}

The complexity analysis of our algorithms will be expressed in terms of the binary complexity of classical operations from computer algebra and algorithmic number theory. In doing so, we have to consider the height of any rational that appears as an input. The notion of height is defined as follows.

\begin{defn}
For $p,q \in \mathbb{Z},$ $q \neq 0$, we define
\[\height(p/q)= \max \left( \lg \vert p \vert , \lg \vert q \vert \right),\]
where $\lg(\cdot)=\frac{\log (\cdot)}{ \log(2)}.$
\end{defn}

We will also use the following definition
to estimate our complexities.

\begin{defn}
For an integer $k>0$ and a map $\phi \: : \: \mathbb{N}^k \rightarrow \mathbb{R}$, we write $\phi (x_1,\dots,x_k)=\poly (x_1,\dots,x_k)$ if there is a polynomial $P \in \mathbb{Q}[x_1,\dots,x_k]$ such that $\phi (x_1,\dots,x_k)=O\left(P(x_1,\dots,x_k)\right).$
\end{defn} 

We now present the complexity of the fundamental operations that we will use.

\begin{defn}
We define $\IntFact(H)$ to be the binary cost of 
factoring an integer $n$ of height $\leq H.$
\end{defn}
Using classical algorithms, such as continued fractions or a general number field sieve, $\IntFact(H)$ is (sub)-exponential in $H$.

\begin{defn}
We define $\PolyFact(d,H)$ to be the binary cost of 
 factoring a polynomial in $\QQx$ of degree $d$ whose coefficients are of heights $\leq H.$

\end{defn}
Using classical algorithms (see, e.g., $\S$ 21 of \cite{AECF17}), $\PolyFact(d,H)$ is polynomial in $d$ and $H$.

\begin{defn}
We define $\UGComp(L,f)$ to be the binary cost of  computing a presentation of the unit group  of the number field $L= \quo{\QQx}{\ideal{f}}$ defined by the irreducible polynomial $f$.
We then define $\UGComp(d,H)$ to be an upper bound on the binary cost of 
 computing a presentation of the unit group
 of a number field~$L$ defined as a quotient $\quo{\QQx}{\ideal{f}}$ with an irreducible polynomial~$f$ of degree~$d$ with coefficients
 of heights $\leq H$. 
 \end{defn} 
 This operation is a central and delicate task in algorithmic
 number theory.
The state-of-the-art algorithms claim sub-exponential complexities
under special assumptions or heuristics (see \cite{biasse2014,  gelin2017, gelin_joux_2016}).
A polynomial quantum algorithm was developed in \cite{biasse2016}.

\begin{defn}
We define $\SFUnits(L,f)$ to be the binary size of the representation
using polynomials in $\QQx$ of
 a system of fundamental units of the unit group of 
 of the number field $L= \quo{\QQx}{\ideal{f}}$ defined
 by the irreducible polynomial $f$ and computed in $\UGComp(L,f)$
 binary complexity.
  $\SFUnits(d,H)$ is defined to be an upper bound on the binary size of 
  the $\SFUnits(L,f)$ for 
  irreducible polynomials $f$ of degree $d$ with coefficients of heights $\leq H$,
  and $L=\quo{\QQx}{\ideal{f}}$. They are computed in $\UGComp(d,H)$
 binary complexity.
 \if{
We define $\SFUnits(d,h)$ to be an upper bound on the binary size of 
the representation
using polynomials in $\QQx$ of
 a system of fundamental units of the unit group of 
 of a number field $L$ defined as a quotient $\quo{\QQx}{\ideal{f}}$ with an irreducible polynomial $f$ of degree $d$ with coefficients
 of heights $\leq h$, and computed in $\UGComp(d,h)$
 binary complexity.
 }\fi
 \end{defn} 
While special compact representations of fundamental units exist (see \cite{thiel1994,thiel1995}), bounding $\SFUnits(L,f)$ or $\SFUnits(d,H)$ is also a delicate  question, depending on the discriminant and regulator of $L$. 
Taking representation with polynomials in $\QQx$, we can only assume that $\SFUnits(d,H)$ is exponential in $d$ and $H$.
\begin{rem}
As all these complexities are (sub)-exponential or polynomial
with a high degree, we can safely assume that
$\PolyFact(d,H)$, $\UGComp(d,H)$, $\SFUnits(d,H)$ are
superlinear in $d$ and $h$,
e.g., $\PolyFact(d,H)+\PolyFact(d',H) \leq \PolyFact(d+d',H)$ for any positive $d,d',H$.
\end{rem}

\section{Solving sums of two squares}
\label{sec:twosquares}
The following fact is well known. It is a special case of \cite[Theorem~17.4]{Rajwade1993}. We present it here for the sake of completeness as it provides an explicit, algorithmic method of deciding whether a given polynomial can be expressed as a sum of two squares.

\begin{obs}\label{obs:sos2}
A square-free polynomial~$f$ is a sum of two squares if and only if the following two conditions hold simultaneously:
\begin{enumerate}
\item $\lc(f)$ is a sum of two squares \textup(in $\QQ$\textup),
\item  $-1$ is a square in $\quo{\QQx}{\ideal{p}}$ for every irreducible factor~$p$ of~$f$.
\end{enumerate}
\end{obs}

We are now ready to present an algorithm (see  Algorithm \ref{alg:sos2}) that decomposes a given polynomial into a sum of two squares.

%\begin{prop}
%Let $f \in \QQx$ be a polynomial of degree $d$ with coefficients
%of heights upper-bounded by $h$, and which can be decomposed as
%$f=a^2+b^2.$
%Then 
%there is some $H=\poly (d,h)$
%such that 
%Algorithm \ref{alg:sos2} computes such $a$ and $b$
%in
%$\IntFact (h)+d\PolyFact(2d,H)$
%binary operations, with output $a,b$
%of heights in $\poly (d,h).$
%\end{prop}

\begin{prop}
Let $f\in \QQx$ be a polynomial which is a sum of two squares. Then Algorithm~\ref{alg:sos2} outputs polynomials $a, b\in \QQx$ such that $a^2 + b^2 = f$.

Moreover, if the degree of~$f$ is~$d$ and the heights of its coefficients are bounded from above by~$H$, some $H_1=\poly (d,H)$, the heights of the output are in $O(H_1)$ and the number of binary operations is upper-bounded by $\IntFact (H)+d\PolyFact(2d,H_1)$. 
\end{prop}

%\begin{alg}\label{alg:sos2}
%Given a polynomial~$f$, which is a priori known to be a sum of two squares in~$\QQx$, this algorithm outputs two polynomials $a, b$ such that $a^2 + b^2 = f$.
%\begin{enumerate}
%\item Construct the quadratic field extension $\QQ(i)/\QQ$.
%\item\label{st:sos2:norm} Solve the norm equation 
%\[
%\lc(f) = \norm[\QQ(i)/\QQ](x)
%\]
%and denote a solution by $a + bi\in \QQ(i)$.
%\item Factor $f$ into a product of monic irreducible polynomials 
%\[
%f = \lc(f)\cdot p_1^{e_1}\dotsm p_k^{e_k}.
%\]
%\item\label{st:sos2:loop} Repeat the following steps for every irreducible factor~$p_j$, such that the corresponding exponent $e_j$ is odd:
%    \begin{enumerate}
%    \item Factor $p_j$ over $\QQ(i)$ into a product $p_j = g_j\cdot h_j$ with $g_j, h_j\in \QQ(i)[x]$.
%    \item Set
%    \[
%    a_j := \frac12\cdot (g_j + h_j),
%    \qquad
%    b_j := \frac{1}{2i}\cdot (g_j - h_j).
%    \]
%    \item Update $a$ and $b$ setting:
%    \[
%    a := aa_j + bb_j
%    \qquad\text{and}\qquad
%    b := ab_j - ba_j.
%    \]
%    \end{enumerate}
%\item Update the polynomials $a$ and $b$ setting
%\[
%a := a\cdot \prod_{j\leq k} p_j^{2 \lfloor \sfrac{e_j}{2}\rfloor}
%\qquad\text{and}\qquad
%b := b\cdot \prod_{j\leq k} p_j^{2 \lfloor \sfrac{e_j}{2}\rfloor}.
%\]
%\item Output $a$ and $b$.
%\end{enumerate}
%\end{alg}

\begin{algorithm}
	\caption{Computing a decomposition of a polynomial as a sum of two squares}
	\label{alg:sos2}
	\begin{algorithmic}[1]
		\REQUIRE A polynomial~$f \in \QQx$, which is a priori known to be a sum of two squares in~$\QQx$.
		\ENSURE Polynomials $a, b\in \QQx$ such that $a^2 + b^2 = f$.
		\STATE Construct the quadratic field extension $\QQ(i)/\QQ$.
        \STATE  Solve the norm equation 
\[
\lc(f) = \norm[\QQ(i)/\QQ](x)
\]
and denote a solution by $a + bi\in \QQ(i)$. \label{st:sos2:norm}
        \STATE Factor $f$ into a product of monic irreducible polynomials 
\[
f = \lc(f)\cdot p_1^{e_1}\dotsm p_k^{e_k}.
\]\label{st:sos2:factor}
		\FOR{ every factor~$p_j$, such that the corresponding exponent $e_j$ is odd \label{st:sos2:loop}}
			\STATE Factor $p_j$ over $\QQ(i)$ into a product $p_j = g_j\cdot h_j$ with $g_j, h_j\in \QQ(i)[x]$.\label{st:sos2:factor_over_Qi}
			\STATE Set
    \[
    a_j := \frac12\cdot (g_j + h_j),
    \qquad
    b_j := \frac{1}{2i}\cdot (g_j - h_j).
    \]
            \STATE Update $a$ and $b$ setting:
    \[
    a := aa_j + bb_j
    \qquad\text{and}\qquad
    b := ab_j - ba_j.
    \]
		\ENDFOR
        \STATE Update $a$ and $b$ setting:
\[
a := a\cdot \prod_{j\leq k} p_j^{2 \lfloor \sfrac{e_j}{2}\rfloor}
\qquad\text{and}\qquad
b := b\cdot \prod_{j\leq k} p_j^{2 \lfloor \sfrac{e_j}{2}\rfloor}.
\]
	\RETURN $a,b$.
	\end{algorithmic}
\end{algorithm}

\begin{poc}[Algorithm \ref{alg:sos2}]
Let $\xi = a + bi \in \QQ(i)$ be the element constructed in line~\ref{st:sos2:norm} of the algorithm. Then 
\[
a^2 + b^2 = \norm[\QQ(i)/\QQ](\xi) = \lc(f)
\]
is a decomposition of the leading coefficient of~$f$ into a sum of two squares. By the previous observation, such decomposition exists. Hence the norm equation has a solution.

Assume now that $p_j$ is a monic irreducible factor of~$f$ of odd multiplicity. The preceding observation asserts that $-1$ is a square in $\quo{\QQx}{\ideal{p_j}}$. 
In other words, the place of $\QQ(x)$ associated to~$p_j$ splits in $\QQ(i, x)$. 
Indeed, since $-1$ is a square in $\quo{\QQx}{\ideal{p_j}}$, we have $-1 = s^2 + q p_j$ for some $q,s \in \QQx$, or equivalently $q p_j = s^2 + 1 = (s-i)(s+i)$. 
The decomposition of $s+i$ into irreducible factors on $\QQ(i)[x]$ does not involve any real factor, only the complex ones since $s+i$ has no real roots. Therefore, the decomposition of $p_j$ in the same ring involves only non-real complex factors (arising from $s+i$) together with their conjugates. 
This implies that $p_j$ factors in $\QQ(i)[x]$ into a product of two conjugate irreducible polynomials $p_j = g_j\cdot h_j$, where
\[
g_j = a_j + b_ji
\qquad\text{and}\qquad
h_j = a_j - b_ji
\]
for some polynomials $a_j, b_j\in \QQx$. It is now clear that $p_j = a_j^2 + b_j^2$ and
\[
a_j = \frac12\cdot (g_j + h_j),
\qquad
b_j = \frac{1}{2i} (g_j - h_j).
\]

Suppose for a moment that $f$ is square-free, i.e., all the exponents~$e_j$ are equal to~$1$. A simple induction combined with the following well-known formula for the product of two sums of squares
\[
(A^2 + B^2)\cdot (a^2 + b^2) =
(Aa + Bb)^2 + (Ab - Ba)^2
\]
shows that after exceeding the loop in line~\ref{st:sos2:loop}, the polynomials $a$ and $b$ satisfy the condition $a^2 + b^2 = f$.

Finally, let $f$ be arbitrary. We can write it as a product $f = gh^2$, where
\[
g := \lc(f)\cdot \prod_{\substack{j\leq k\\ 2\nmid e_j}} p_j
\qquad\text{and}\qquad
h := \prod_{j\leq k} p_j^{\lfloor\sfrac{e_j}{2}\rfloor}.
\]
Then $g$ is square-free and decomposes into a sum of two squares $g = a^2 + b^2$ by the preceding part of the proof. It follows that $f = (ah)^2 + (bh)^2$, and this proves the correctness of the algorithm.
\end{poc}

\begin{comp_analysis}[Algorithm \ref{alg:sos2}]
Let $f$ be of degree $d$ and with coefficients
of heights upper-bounded by $H$.
To solve the norm equation in line~\ref{st:sos2:norm},
it is enough to: \begin{inparaenum} \item factor $\lc(f)$, \item solve the norm equation for each prime factor, \item combine the solutions using the Brahmagupta–Fibonacci identity, i.e., $(\alpha^2+\beta^2)(\gamma^2+\delta^2)=(\alpha\gamma-\beta\delta)^2+(\alpha\delta+\beta\gamma)^2=(\alpha\gamma+\beta\delta)^2+(\alpha\delta-\beta\gamma)^2$.  
\end{inparaenum}
The first item is in $\IntFact (H)$.
Thanks to \cite[p. 128]{Wagon90} and \cite{Schoof85},
the second item is polynomial in $H$. 
The recombination is also polynomial in $H$
and the output solutions are of heights upper-bounded
by $H$. 

Line~\ref{st:sos2:factor} is in $\PolyFact(n,H)$.
Line~\ref{st:sos2:factor_over_Qi} is done in $\PolyFact(2d,H_1)$
for some $H_1$ polynomial in $H$ and $d$, by \cite[\S~3.6.2]{Cohen93}.
After that step, there are only arithmetic operations.

All in all, for some $H_1$ polynomial in $H$ and $d$,
the total complexity of Algorithm \ref{alg:sos2}
is in $\IntFact (H)+d\PolyFact(2d,H_1)$.
\end{comp_analysis}

\begin{rem}
To solve the norm equation in line~\ref{st:sos2:norm} of Algorithm~\ref{alg:sos2} (as well as in line~\ref{st:initial:norm} of Algorithm~\ref{alg:initial} below), it is possible to rely also on the methods described in \cite{Cohen2000, FJP1997, FP1983, Garbanati1980, Simon2002}.
\end{rem}

\section{Solving sums of four squares}
\label{sec:foursquares}
The goal of this section is to design and analyze a procedure to decompose a nonnegative element of $\QQ[x]$ into a sum of four squares in $\QQ[x]$, assuming that such a decomposition exists. 
This is achieved via Algorithm~\ref{alg:initial} and Algorithm~\ref{alg:irreducible_sos4} below and requires computing a decomposition of $-1$ into a sum of two squares in the number field~$K:= \quo{\QQx}{\ideal{f}}$. 
This basically boils down to solving a norm equation $-1 = \norm{x}$, where $L = K\bigl(\sqrt{-1}\bigr)$. 
To this end, one can use any of the methods described in \cite{Cohen2000, FJP1997, FP1983, Garbanati1980, Simon2002}. 
Nevertheless, to make it possible to perform the complexity analysis of Algorithm~\ref{alg:initial}, we provide in Algorithm~\ref{alg:norm_eq} a 
stripped-down version of Simon's algorithm for solving a norm equation (cf. \cite{Simon2002}) which suffices to our purpose. 

Recall that a field~$K$ is called \term{non-real} if $-1$ is a sum of squares in~$K$. The minimal number of summands needed to express $-1$ as a sum of squares is called the \term{level} of~$K$ and denoted $s(K)$.
%
%\begin{alg}\label{alg:norm_eq}
%Given a number field~$K$ such that $-1$ is a sum of two squares in~$K$, this algorithm outputs $a,b\in K$ satisfying $a^2 + b^2 = -1$.
%\begin{enumerate}
%\item\label{st:norm_eq:level_1} If there is $c\in K$ such that $c^2 = -1$, output $a = c$, $b = 0$ and quit.
%\item Construct the quadratic field extension $L := K\bigl(\sqrt{-1}\bigr)$.
%\item Construct the unit groups $U_K$ and $U_L$ of~$K$ and $L$, respectively.
%\item Construct the quotient groups $G_K := \quo{U_K}{U_K^2}$ and $G_L := \quo{U_L}{U_L^2}$. Let $\SK := \{ \kappa_1, \dotsc, \kappa_k \}\subset U_K$ and $\SL := \{ \lambda_1, \dotsc, \lambda_l \} \subset U_L$ be sets of elements forming bases of~$G_K$ and $G_L$, treated as $\FF_2$-vector spaces.
%\item\label{st:norm_eq:V} Let $V = (v_1, \dotsc, v_k)$ be the coordinates with respect to~$\SK$ of the coset $-1\cdot U_K^2$.
%\item\label{st:norm_eq:M} For every $i\leq l$ denote the coordinates of $\norm(\lambda_i)\cdot U_K^2$ with respect to~$\SK$ by $(m_{i1},\dotsc, m_{ik})$.
%\item\label{st:norm_eq:solve} Solve the system of $\FF_2$-liner equations $M^T\cdot X = V$, where $M = (m_{ij})$. Denote a solution by $(\varepsilon_1, \dotsc, \varepsilon_l)$.
%\item Set $\lambda := \lambda_1^{\varepsilon_1}\dotsm \lambda_l^{\varepsilon_l}$ and let $c \in K$ be such that $c^2 = -\norm(\lambda)$.
%\item  Write $\sfrac{\lambda}{c}$ as $\sfrac{\lambda}{c} = a + b\sqrt{-1}$ and output $a$ and $b$.
%\end{enumerate}
%\end{alg}

\begin{algorithm}
	\caption{Decomposing $-1$ as a sum of two squares over the number field $K:= \quo{\QQx}{\ideal{f}}$.}
	\label{alg:norm_eq}
	\begin{algorithmic}[1]
		\REQUIRE Number field $K$ of level $s(K)\leq 2$.
		\ENSURE $a, b\in K$ such that $a^2 + b^2 = -1$.
		\IF{there is $c\in K$ such that $c^2 = -1$ \label{st:norm_eq:level_1}}
            \RETURN $a = c$, $b = 0$.
        \ENDIF 
        \STATE  Construct the quadratic field extension $L := K\bigl(\sqrt{-1}\bigr)$.
        \STATE Construct the unit groups $U_K$ and $U_L$ of~$K$ and $L$, respectively. \label{st:unit_group_computation}
        \STATE Construct the quotient groups $G_K := \quo{U_K}{U_K^2}$ and $G_L := \quo{U_L}{U_L^2}$. Let $\SK := \{ \kappa_1, \dotsc, \kappa_k \}\subset U_K$ and $\SL := \{ \lambda_1, \dotsc, \lambda_l \} \subset U_L$ be sets of elements forming bases of~$G_K$ and $G_L$, treated as $\FF_2$-vector spaces.
        \STATE Let $V = (v_1, \dotsc, v_k)$ be the coordinates with respect to~$\SK$ of the coset $-1\cdot U_K^2$. \label{st:norm_eq:V}
        \STATE For every $i\leq l$ denote by $(m_{i1},\dotsc, m_{ik})$ the coordinates of $\norm(\lambda_i)\cdot U_K^2$ with respect to~$\SK$. \label{st:norm_eq:M}
        \STATE Solve the system of $\FF_2$-liner equations $M^T\cdot X = V$, where $M = (m_{ij})$. Denote a solution by $(\varepsilon_1, \dotsc, \varepsilon_l)$. \label{st:norm_eq:solve} 
        \STATE Set $\lambda := \lambda_1^{\varepsilon_1}\dotsm \lambda_l^{\varepsilon_l}$ and let $c \in K$ be such that $c^2 = -\norm(\lambda)$.
        \STATE Write $\sfrac{\lambda}{c}$ as $\sfrac{\lambda}{c} = a + b\sqrt{-1}$. 
		\RETURN $a,b$.
	\end{algorithmic}
\end{algorithm}

%The above algorithm is nothing else but a stripped-down version of Simon's algorithm for solving a norm equation (cf. \cite{Simon2002}).  just enough to present a complexity analysis of Algorithm~\ref{alg:initial}. 
%Due to the fact that it is a variant of a well-known algorithm, we feel free to omit the proof of correctness.

%\todo[inline]{
%I believe that in the final version of the paper we can freely omit the proof of correctness of Algorithm~\ref{alg:norm_eq}, but for our own sake I include below a detailed proof, so we know for sure that there is not gap in the argumentation}

%\begin{prop}
%Let $f \in \QQx$ be a polynomial of degree $d$ with coefficients
%of heights upper-bounded by $h$, and such that $-1=a^2+b^2$
%for some $a,b \in \quo{\QQx}{\ideal{f}}$.
%Then 
%Algorithm \ref{alg:norm_eq} computes such $a$ and $b$
%in binary complexity polynomial in 
%$\UGComp (2d,2hd+8d(1+\lg (d)))$
%and $\SFUnits(2d,2hd+8d(1+\lg (d)))$.
%The output $a,b$
%have binary-size polynomial in
%$\SFUnits(2d,2hd+8d(1+\lg (d)))$.
%\end{prop} 

\begin{prop}
Let $K = \quo{\QQx}{\ideal{f}}$ be a number field of level $s(K)\leq 2$ and specified by its generating polynomial~$f$. Then Algorithm~\ref{alg:norm_eq} outputs $a, b\in K$ such that $a^2 + b^2 = -1$.

Moreover, if the degree of~$f$ is $d$ and the heights of its coefficients are bounded from above by~$H$, then $a,b$ have binary-size polynomial in $\SFUnits(2d,2Hd+8d(1+\lg (d)))$ and the number of binary operations is polynomial in $\UGComp (2d,2Hd+8d(1+\lg (d)))$ and $\SFUnits(2d,2Hd+8d(1+\lg (d)))$.
\end{prop}
 
\begin{poc}[Algorithm \ref{alg:norm_eq}]
If the algorithm terminates in line~\ref{st:norm_eq:level_1}, then the correctness of its output is trivial. Hence, without loss of generality, we may assume that $-1\notin K^{\times 2}$. In particular, this means that $L = K\bigl(\sqrt{-1}\bigr)$ is a proper quadratic extension of~$K$. By assumption $-1$ is a sum of two squares in~$K$, say $-1 = a^2 + b^2$ with $a, b\in K^\times$. Then $-1 = \norm(\mu)$, where $\mu = a + b\sqrt{-1}$. Observe that $\mu$ must be a unit in~$L$ since 
\[
\norm[\quo{L}{\QQ}](\mu) 
= ( \norm[\quo{K}{\QQ}]\circ \norm )(\mu)
= \norm[\quo{K}{\QQ}](-1)
= (-1)^{(K:\QQ)}.
\]
It follows that we can write the element~$\mu$ as 
\[
\mu = \lambda_1^{\varepsilon_1}\dotsm \lambda_l^{\varepsilon_l}\cdot \nu^2,
\]
where $\nu\in U_L$ and $\varepsilon_1, \dotsc, \varepsilon_l\in \{0, 1\}$ are the coordinates of the coset $\mu\cdot U_L^2$ with respect to the basis $\SL = \{\lambda_1, \dotsc, \lambda_l \}$ of $G_l = \quo{U_L}{U_L^2}$. Computing the norms of both sides of the above equation we obtain 
\[
-1
= \norm(\mu)
= \norm(\lambda_1)^{\varepsilon_1}\dotsm \norm(\lambda_l)^{\varepsilon_l}\cdot \norm(\nu)^2.
\]
Now, let $m_{ij}$ and $v_j$ with $i\leq l$, $j\leq k$ be as in steps (\ref{st:norm_eq:V}--\ref{st:norm_eq:M}) of the algorithm. In the quotient group $G_K = \quo{U_K}{U_K^2}$ we have 
\[
\bigl( \kappa_1^{v_1}\dotsm \kappa_k^{v_k} \bigr)\cdot U_K^2
= 
\prod_{i\leq l}\prod_{j\leq k} \kappa_j^{m_{ij}\varepsilon_i} \cdot U_K^2.
\]
The cosets of $\kappa_1, \dotsc, \kappa_k$ form a basis of~$G_K$. In particular they are linearly independent. Therefore for every $j\leq k$ we have
\[
v_j = m_{1j}\varepsilon_1 + \dotsb + m_{lj}\varepsilon_l.
\]
This way we have proved that the system of $\FF_2$-linear equations considered in line~\ref{st:norm_eq:solve} has a solution. 

Conversely, let $\varepsilon_1, \dotsc, \varepsilon_l\in \{0,1\}$ form a solution to the system $M^T\cdot X = V$ and let $\lambda = \lambda_1^{\varepsilon_1}\dotsm \lambda_l^{\varepsilon_l}$. It follows from the preceding part of the proof that the cosets $-1\cdot U_K^2$ and $\norm(\lambda)\cdot U_K^2$ coincide. Therefore, there is a unit $c\in U_K$ such that $-c^2 = \norm(\lambda)$. Consequently $-1 = \norm(\sfrac{\lambda}{c})$ and so if $\sfrac{\lambda}{c} = a + b\sqrt{-1}$, we obtain the sought decomposition $-1 = a^2 + b^2$. 
\end{poc}

\begin{comp_analysis}[Algorithm \ref{alg:norm_eq}]
We assume that $K= \quo{\QQx}{\ideal{f}}$ with
$f$ of degree $d$, monic, irreducible, with coefficients of heights $\leq H$. 
Let us first bound the costs of constructing~$L$. 
Let $g=x^2+1$, then
for some $c\in \mathbb{Z}$ with $\vert c \vert \leq d^2$,
$u(x):=\mathrm{Res}_t(f(t),g(x-ct))$
is an irreducible polynomial defining $L.$
Indeed, it is enough that $c$
is not a quotient of a difference of roots of $f$
by roots of $g$, and there are strictly less than
$d^2$ such quotients.
Since $f$ and $g$ are monic,
$u(x)=f (\sfrac{(x-i)}{c}) f( \sfrac{(x+i)}{c} )$.
The heights of the coefficients of $u$
can then be upper-bounded by $2Hd+8d(1+\lg (d))$. 
This implies that the costs of computing $U_K$
and $U_L$ in line~\ref{st:unit_group_computation}
are in $\UGComp (2d,2Hd+8d(1+\lg (d)))$.
The construction of the quotient group from the system of fundamental units is direct.
The linear system in line~\ref{st:norm_eq:solve} is solved in time polynomial in $d$ and $\SFUnits(2d,2Hd+8d(1+\lg (d)))$. 
The output bit-size of $a,b$ is polynomial in
$\SFUnits(2d,2Hd+8d(1+\lg (d)))$.
\end{comp_analysis}

%\begin{alg}[Initial solution]\label{alg:initial}
%Given an irreducible polynomial $f\in \QQx$, which is a priori known to be a sum of~$3$ or~$4$ squares, this algorithm outputs polynomials $h$ and $g_1, \dotsc, g_4$, such that $\deg h\leq \deg f-2$ and $f h = g_1^2 +\dotsb + g_4^2$.
%\begin{enumerate}
%\item Construct the number fields: 
%\[ 
%K := \quo{\QQx}{\ideal{f}}\qquad\text{and}\qquad L := K(i). 
%\]
%\item\label{st:initial:norm} Solve the norm equation
%\[
%-1 = \norm(x)
%\]
%and denote the solution by $\xi = \og_1 + \og_2 i$, where $g_1, g_2\in \QQx$ are polynomials of degree strictly less than $\deg f$ and $\og_j$ denotes the image of~$g_j$ under the canonical epimorphism $\QQx\onto K$. 
%\item Set $g_3 := 1$, $g_4 := 0$ and let $h := \sfrac{(g_1^2 +\dotsb + g_4^2)}{f}$
%\item Output $h, g_1, g_2, g_3, g_4$.
%\end{enumerate}
%\end{alg}
\begin{algorithm}
	\caption{Initial solution: modular sum of squares}
	\label{alg:initial}
	\begin{algorithmic}[1]
		\REQUIRE An irreducible polynomial $f\in \QQx$, which is a priori known to be a sum of~$3$ or~$4$ squares.
		\ENSURE Polynomials $h$ and $g_1, \dotsc, g_4$ in $\QQx$, such that $\deg h\leq \deg f-2$ and $f h = g_1^2 +\dotsb + g_4^2$.
		\STATE  Construct the number fields: 
\[ 
K := \quo{\QQx}{\ideal{f}}\qquad\text{and}\qquad L := K(i). 
\]
        \STATE  Solve the norm equation
\[
-1 = \norm(x)
\]
and denote the solution by $\xi = \og_1 + \og_2 i$, where $g_1, g_2\in \QQx$ are polynomials of degree strictly less than $\deg f$ and $\og_j$ denotes the image of~$g_j$ under the canonical epimorphism $\QQx\onto K$. \label{st:initial:norm}
        \STATE Set $g_3 := 1$, $g_4 := 0$ and let $h := \sfrac{(g_1^2 +\dotsb + g_4^2)}{f}$.
		\RETURN $h, g_1, g_2, g_3, g_4$.
	\end{algorithmic}
\end{algorithm}

\begin{prop}
Let $f \in \QQx$ be an irreducible polynomial of degree $d$ with coefficients
of heights upper-bounded by $H$, and known to be a sum of $3$ or $4$ squares in $\QQx$.
Then 
Algorithm \ref{alg:initial} computes 
polynomials $h$ and $g_1, \dotsc, g_4$ in $\QQx$, 
such that $\deg h\leq \deg f-2$ and $f h = g_1^2 +\dotsb + g_4^2$.

The binary complexity is polynomial 
in $\UGComp (2d,2Hd+8d(1+\lg (d)))$
and $\SFUnits(2d,2Hd+8d(1+\lg (d)))$.
The outputs are of binary size polynomial in
$\SFUnits(2d,2Hd+8d(1+\lg (d)))$.
\end{prop}
\begin{poc}[Algorithm \ref{alg:initial}]
By assumption $f$ is a sum of~$4$ or fewer squares of polynomials, say
\[
f = p_1^2 + p_2^2 + p_3^2 + p_4^2.
\]
Not all $p_j$ are zeros. Assume that $p_1\neq 0$. Then in $K$ we have
\[
-1 
= \left(\frac{\op_2}{\op_1}\right)^2 
+ \left(\frac{\op_3}{\op_1}\right)^2 
+ \left(\frac{\op_4}{\op_1}\right)^2.
\]
Here again, $\op_j$ denotes the image of~$p_j$ in~$K$. It follows that $K$ is a non-real field, and its level does not exceed~$3$. But it is well known (see, e.g., \cite[Theorem~XI.2.2]{Lam2005}) that the level of a non-real field is always a power of~$2$. Therefore, $-1$ is a sum of~$2$ squares in~$K$. Thus, there are polynomials $g_1, g_2\in \QQx$ such that $\deg g_j< \deg f$ and
\[
\og_1^2 + \og_2^2 = \norm(\og_1 + \og_2 i) = -1.
\]
It is now clear that $g_1^2 + g_2^2 + 1$ is divisible by~$f$. Denoting the quotient by~$h$ we obtain $fh = g_1^2 + g_2^2 + 1$. Moreover, we have $\deg g \leq \deg f - 1$ and so $\deg h$ cannot exceed $\deg f - 2$, as expected. 
\end{poc}

\begin{comp_analysis}[Algorithm \ref{alg:initial}]
By the analysis of Algorithm \ref{alg:norm_eq},
if $f$ is of degree $d$ and its coefficients of 
heights $\leq H$, then the binary complexity
of Algorithm \ref{alg:initial}
is polynomial in $\UGComp (2d,2Hd+8d(1+\lg (d)))$
and $\SFUnits(2d,2Hd+8d(1+\lg (d)))$.
The outputs are also of binary size polynomial in
$\SFUnits(2d,2Hd+8d(1+\lg (d)))$.
\end{comp_analysis}

We are now ready to present a method (see Algorithm~\ref{alg:irreducible_sos4}) to decompose an irreducible rational polynomial as a sum of 3 or 4 squares in $\QQx$, provided that such a decomposition exists. 
We need to fix which of the different variants of Euler's identity will be used for multiplying sums of four squares. The following one ensures that the succeeding algorithm will output polynomials.

\begin{EI}
\begin{align*}
\mathrlap{(A^2 + B^2 + C^2 + D^2) (a^2 + b^2 + c^2 + d^2) =}\hspace*{20mm} &\\
&= (Aa + Bb + Cc + Dd)^2 + (-Ab + Ba - Cd + Dc)^2 \\
&\,+ (-Ac + Bd + Ca - Db)^2 + (-Ad - Bc + Cb + Da)^2.
\end{align*}
\end{EI}

\begin{algorithm}
	\caption{Decomposition of a monic irreducible polynomial into a sum
 of 3 or 4 squares}
	\label{alg:irreducible_sos4}
	\begin{algorithmic}[1]
		\REQUIRE A monic irreducible polynomial~$f  \in  \QQx$, which is a priori known to be a sum of~$3$ or $4$ squares.
		\ENSURE Polynomials $f_1, \dotsc, f_4\in  \QQx$ such that $f_1^2 + \dotsb + f_4^2 = f$. 
		\STATE Execute Algorithm~\ref{alg:initial} to construct polynomials $h, g_1, \dotsc, g_4$, such that $fh = g_1^2 + \dotsb + g_4^2$ and $\deg h \leq \deg f - 2$.\label{st:irreducible_sos4:init}
        \WHILE{$\deg h > 0$}
        \STATE Compute the remainders $r_j := (g_j\bmod h)$ of $g_1, \dotsc, g_4$ modulo~$h$.
        \STATE Using Euler's identity, express the product
    \[
    \bigl(g_1^2 + g_2^2 + g_3^2 + g_4^2\bigr)\bigl(r_1^2 + r_2^2 + r_3^2 + r_4^2\bigr)
    \]
    as a sum of four squares. Denote the result by $g'_1, \dotsc, g'_4$. \label{it:irreducible_sos4:Euler}
        \STATE Update $g_1, \dotsc, g_4$ setting $g_j := \sfrac{g'_j}{h}$ for $j\in \{1,2,3,4\}$.
        \STATE Update~$h$ setting $h := \sfrac{(g_1^2 + \dotsb + g_4^2)}{f}$. \label{st:irreducible_sos4:update_h}
        \ENDWHILE
        \STATE Decompose~$h$ into a sum of fours squares of rational numbers $h = a_1^2 + \dotsb + a_4^2$, where $a_1, \dotsc, a_4\in \QQ$. \label{st:irreducible_sos4:QQ}
        \STATE Use Euler's identity to express the product 
\[
\Bigl( g_1^2 +\dotsb + g_4^2 \Bigr) 
\cdot
\left( 
\left(\frac{a_1}{h}\right)^2 + \dotsb + \left(\frac{a_4}{h}\right)^2
\right)
\]
as a sum of four squares. Store the result in $f_1, \dotsc, f_4$. \label{st:irreducible_sos4:form_result}
		\RETURN  $f_1, \dotsc, f_4$.
	\end{algorithmic}
\end{algorithm}

\begin{prop}
Let $f \in \QQx$ be an irreducible polynomial of degree $d$ with coefficients
of heights upper-bounded by $H$, and known to be a sum of $3$ or $4$ squares in $\QQx$.
Then 
Algorithm \ref{alg:irreducible_sos4} computes 
polynomials $g_1, \dotsc, g_4$ in $\QQx$, 
such that $f = g_1^2 +\dotsb + g_4^2$.

The binary complexity is polynomial 
in $\UGComp (2d,2Hd+8d(1+\lg (d)))$
and $\SFUnits(2d,2Hd+8d(1+\lg (d)))$.
The output is of binary size polynomial in
$\SFUnits(2d,2Hd+8d(1+\lg (d)))$.
\end{prop}

\begin{poc}[Algorithm \ref{alg:irreducible_sos4}]
We shall denote the values of the variables $h$, $g_j$, $g'_j$ and~$r_j$ after the $k$-th iteration of the main loop by $h_k$, $g_{kj}$, $g'_{kj}$ and~$r_{kj}$, respectively. In particular, the initial values computed in line~\ref{st:irreducible_sos4:init} will be denoted by $h_0$ and $g_{01}, \dotsc, g_{04}$.

By the means of line~\ref{st:irreducible_sos4:update_h}, for every $k\geq 0$, we have $fh_k = g_{k1}^2 + \dotsb + g_{k4}^2$. We shall prove that all $h_k$'s are polynomials and their degrees form a strictly decreasing sequence and likewise that all $g_{k1}, \dotsc, g_{k4}$ are polynomials, too.  Of course, $h_0$ and $g_{01}, \dotsc, g_{04}$ are polynomials. Let us examine the product of two sums of squares that appears in line~\ref{it:irreducible_sos4:Euler} of the algorithm:
\[
\bigl( (g'_{k+1,1})^2 + \dotsb + (g'_{k+1,4})^2 \bigr)
= \bigl(g_1^2 + g_2^2 + g_3^2 + g_4^2\bigr)\bigl(r_1^2 + r_2^2 + r_3^2 + r_4^2\bigr).
\]
Here for every $j\leq 4$ we have $r_{kj} = g_{kj} - q_{kj}h_k$ for some polynomial $q_{kj}\in \QQx$ and $\deg r_{kj} < \deg h_k$. Euler's identity yields
\begin{align*}
g'_{k+1,j}
 = & \ g_{k1}r_{k1} + g_{k2}r_{k2} + g_{k3}r_{k3} + g_{k4}r_{k4} 
\\
 = & \ g_{k1}\cdot (g_{k1} - q_{k1}h_k) + g_{k2}\cdot (g_{k2} - q_{k2}h_k) \\
     & + g_{k3}\cdot (g_{k3} - q_{k3}h_k) + g_{k3}\cdot (g_{k3} - q_{k3}h_k) \\
= & \ \bigl( g_{k1}^2 + g_{k2}^2 + g_{k3}^2 + g_{k4}^2 \bigr)\\
   & - \bigl( g_{k1}q_{k1} + g_{k2}q_{k2} + g_{k3}q_{k3} + g_{k4}q_{k4} \bigr)\cdot h_k \\
= & \ \bigl( f - g_{k1}q_{k1} - g_{k2}q_{k2} - g_{k3}q_{k3} - g_{k4}q_{k4} \bigr)\cdot h_k.
\end{align*}
This shows that $g'_{k+1,1}$ is divisible by~$h_k$, hence $g_{k+1,1} = \sfrac{g'_{k+1,1}}{h_k}$ is indeed a polynomial. Likewise
\begin{align*}
g'_{k+1,2}
= & \ -g_{k1}r_{k2} + g_{k2}r_{k1} - g_{k3}r_{k4} + g_{k4}r_{k3} 
\\
= & \ - g_{k1}\cdot ( g_{k2} - q_{k2}h_k) + g_{k2}\cdot ( g_{k1} - q_{k1}h_k) \\
    & - g_{k3}\cdot ( g_{k4} - q_{k4}h_k) + g_{k4}\cdot ( g_{k3} - q_{k3}h_k) 
\\
= & \ ( g_{k1}q_{k2} - g_{k2}q_{k1} + g_{k3}q_{k4} - g_{k4}q_{k3} )\cdot h_k.
\end{align*}
Therefore, $g_{k+1,2} = \sfrac{g'_{k+1,2}}{h_k}$ is a polynomial, too. Analogous arguments apply to $g_{k+1,3}$ and $g_{k+1,4}$, as well.

Now, assume that we have proved that $h_k$ is a polynomial for some $k\geq 0$. We have
\begin{equation}\label{eq:sum_of_gs}
fh_k = g_{k1}^2 + \dotsb + g_{k4}^2
\end{equation}
and $r_{kj} = (g_{kj}\bmod h_k)$. We deduce that 
\[
r_{k1}^2 + \dotsb + r_{k4}^2 \equiv 0\pmod{h_k}.
\]
Hence, there exists some $h_k'$ such that
\begin{equation}\label{eq:sum_of_rs}
h_kh_k' = r_{k1}^2 + \dotsb + r_{k4}^2.
\end{equation}
Now, $\deg r_{kj} < \deg h_k$ for all $j\leq 4$ and so $\deg h'_k \leq \deg h_k - 2$. Combining \eqref{eq:sum_of_gs} with \eqref{eq:sum_of_rs} we obtain
\[
f\cdot h_k^2\cdot h'_k
= (g_{k1}^2 + \dotsb + g_{k4}^2)(r_{k1}^2 + \dotsb + t_{k4}^2)
= (g'_{k+1,1})^2 + \dotsb + (g'_{k+1,k})^2.
\]
This yields $fh'_k = g_{k+1,1}^2 + \dotsb + g_{k+1,4}^2$. It follows that
\[
h'_k = \frac{ g_{k+1,1}^2 + \dotsb + g_{k+1,4}^2 }{ f } = h_{k+1}. 
\]
It shows that $h_{k+1}$ is a polynomial of degree $\leq \deg h_k - 2$, proving our claim. Consequently, after finitely many steps the degree of~$h_k$ will eventually drop to zero and so the algorithm will terminate.

Now, let $h_k$ be such that $\deg h_k = 0$. We know that
\[
fh_k = g_{k1}^2 + \dotsb + g_{k4}^2.
\]
By assumption, $f$ itself is a sum of squares. Therefore $h_k$ must be a positive rational number, hence a sum of four (or fewer) squares. Write $h_k = a_1^2 + \dotsb + a_4^2$ for some rational numbers $a_1, \dotsc, a_4\in \QQ$. It follows that the product in line~\ref{st:irreducible_sos4:form_result} of the algorithm is a sum of four squares that equals~$f$. This proves the correctness of the presented algorithm.
\end{poc}

\begin{comp_analysis}[Algorithm \ref{alg:irreducible_sos4}]
Thanks to the analysis of Algorithm \ref{alg:initial},
if the polynomial~$f$ is of degree~$d$ and the heights of its coefficients are bounded from above by~$H$,
then 
line~\ref{st:irreducible_sos4:init} has a binary
complexity polynomial in $\UGComp (2d,2Hd+8d(1+\lg (d)))$
and $\SFUnits(2d,2Hd+8d(1+\lg (d)))$.
The outputs are of binary size polynomial in
$\SFUnits(2d,2Hd+8d(1+\lg (d)))$.
The \textbf{while} will be executed less than
$d$-times
and thus involves only a polynomial in $d$
amount of arithmetic operations.

Line~\ref{st:irreducible_sos4:QQ} has an
expected (Las Vegas) binary complexity 
polynomial in the height of $h$, by \cite{RS1986,PT2018}.
All in all, the algorithm has an expected binary
complexity polynomial in $\UGComp (2d,2Hd+8d(1+\lg (d)))$
and $\SFUnits(2d,2Hd+8d(1+\lg (d)))$
and the outputs are still of binary size polynomial in
$\SFUnits(2d,2Hd+8d(1+\lg (d)))$.
\end{comp_analysis}

A decomposition of an arbitrary polynomial into a sum of four squares (provided that such decomposition exists) is now straightforward. For the sake of completeness, we present an explicit algorithm below.

%\begin{alg}\label{alg:sos4}
%Given a polynomial $f\in \QQx$, which is a priori known to be a sum of~$3$ or~$4$ squares, this algorithm outputs polynomials $f_1, \dotsc, f_4$ such that $f_1^2 + \dotsb + f_4^2 = f$.
%\begin{enumerate}
%\item Using square-free decomposition, factor $f$ into a product $f = \lc(f)\cdot g\cdot h^2$, where $g$ is monic and square-free.
%\item\label{st:sos4:QQ} Decompose the leading coefficient of~$f$ into a sum of four squares of rational numbers and initialize $f_1, \dotsc, f_4$ with the result.
%\item Factor $g$ into a product of monic irredu­cible polynomials: $g = p_1\dotsm p_k$.
%\item Repeat the following steps for each $p_j$. 
%    \begin{enumerate}
%    \item Execute Algorithm~\ref{alg:irreducible_sos4} to express $p_j$ as a sum of~$4$ squares
%    \[
%    p_j = g_1^2 + \dotsb + g_4^2.
%    \]
%    \item Use Euler's identity to express the product $(f_1^2 +\dotsb + f_4^2) (g_1^2 +\dotsb + g_4^2)$ as a sum of~$4$ squares. Store the result again in $f_1, \dotsc, f_4$.
%    \end{enumerate}
%\item Output $f_1, \dotsc, f_4$.
%\end{enumerate}
%\end{alg}

\begin{algorithm}
	\caption{Decomposition into a sum  of 3 or 4 squares}
	\label{alg:sos4}
	\begin{algorithmic}[1]
		\REQUIRE A polynomial~$f  \in  \QQx$, which is a priori known to be a sum of~$3$ or $4$ squares.
		\ENSURE Polynomials $f_1, \dotsc, f_4\in  \QQx$ such that $f_1^2 + \dotsb + f_4^2 = f$. 
		\STATE Using square-free decomposition, construct $g,h \in \QQx$ such that $f = \lc(f)\cdot g\cdot h^2$, where $g$ is monic and square-free.
        \STATE Decompose $\lc(f)$ into a sum of four squares of rational numbers and initialize $f_1, \dotsc, f_4$ with the result. \label{st:sos4:QQ}
        \STATE Factor $g$ into a product of monic irreducible polynomials: $g = p_1\dotsm p_k$.
        \FOR{each $p_j$}
        \STATE Execute Algorithm~\ref{alg:irreducible_sos4} to express $p_j$ as a sum of~$4$ squares
    \[
    p_j = g_1^2 + \dotsb + g_4^2.
    \]
        \STATE Use Euler's identity to express the product $(f_1^2 +\dotsb + f_4^2) (g_1^2 +\dotsb + g_4^2)$ as a sum of~$4$ squares. Store the result again in $f_1, \dotsc, f_4$.
        \ENDFOR
		\RETURN  $f_1, \dotsc, f_4$.
	\end{algorithmic}
\end{algorithm}

\begin{prop}
Let $f \in \QQx$ be a polynomial of degree $d$ with coefficients
of heights upper-bounded by $H$, and known to be a sum of $3$ or $4$ squares.
Then 
Algorithm \ref{alg:sos4} computes 
polynomials $g_1, \dotsc, g_4$ in $\QQx$, 
such that $f = g_1^2 +\dotsb + g_4^2$.\\
The expected binary complexity is polynomial 
in $\UGComp (2d,2Hd+8d(1+\lg (d)))$
and $\SFUnits(2d,2Hd+8d(1+\lg (d)))$.\\
The output is of binary size polynomial in
$\SFUnits(2d,2Hd+8d(1+\lg (d)))$.
\end{prop}

In view of Algorithm~\ref{alg:irreducible_sos4}, the correctness of Algorithm~\ref{alg:sos4} is obvious. Hence, we restrict ourselves to the complexity analysis.

\begin{comp_analysis}[Algorithm \ref{alg:sos4}]
The algorithm relies on polynomial factorization,
application of Algorithm~\ref{alg:irreducible_sos4}
for some factors and recombination using
Euler's identity.

Thanks to our super-linearity assumption,
we can upper-bound the sum of the costs of, e.g.,  the $\UGComp$'s of the factors
by $\UGComp (2d,2Hd+8d(1+\lg (d)))$.
All in all, the algorithm has an expected binary
complexity polynomial in $\PolyFact (d,H),$ $\UGComp (2d,2Hd+8d(1+\lg (d)))$
and $\SFUnits(2d,2Hd+8d(1+\lg (d)))$
and the outputs are still of binary size polynomial in
$\SFUnits(2d,2Hd+8d(1+\lg (d)))$.
\end{comp_analysis}

\begin{rem}
In order to decompose a natural number into a sum of four squares (see line~\ref{st:irreducible_sos4:QQ} of Algorithm~\ref{alg:irreducible_sos4} and line~\ref{st:sos4:QQ} of Algorithm~\ref{alg:sos4}) one can use, e.g., \cite{Bumby1996, PS2019, PT2018, RS1986}.
If one needs a deterministic algorithm, one can \begin{inparaenum}    \item Factor the entry $z$ over $\mathbb{Z}$,
    \item For each prime factor $p$, compute a decomposition of $p$ as 
    a sum of four squares using \cite[\S 4.3]{Bumby1996} and then \cite[\S 3]{RS1986}
    \item Recombine using Euler's identity.
\end{inparaenum}
The complexity is then polynomial in $\IntFact (\height(z)),$ and $\height(z)$.
\end{rem}

\section{Reduction to four squares}
\label{sec:fivesquares}

In this Section we introduce Algorithms \ref{alg:reduce_helper} and \ref{alg:reduce} that reduce
our main problem of decomposing a positive polynomial
into a sum of squares to the problem of decomposing a polynomial
into a sum of four squares, which we already tackled
in the previous section.
Given $f = c_0 + c_1x + \dotsb + c_d x^d\in \QQx$, we define the associated reciprocal polynomial $f_* := c_d + c_{d-1} x + \dotsb + c_0 x^d\in \QQx$.
First we focus on the case where the 2-adic valuation of $c_d$ is odd, in which case we rely on Algorithm \ref{alg:reduce_helper}, and next we handle the general case via Algorithm \ref{alg:reduce}.

%\todo[inline]{Tristan: I have added "not a sum of 4 squares" as assumption on the input of Algo \ref{alg:reduce_helper} and \ref{alg:reduce}
%since, unless I am mistaken, when they are called in Algo \ref{alg:sixsquares}, it has already been checked.}

%\begin{alg}\label{alg:reduce_helper}
%Let $f = f_0 + f_1x + \dotsb + f_dx^d\in \QQx$ be a positive square-free polynomial. Denote the $2$-adic valuations of the coefficients of~$f$ by $k_j := \ord_2 f_j$ for $0\leq j\leq d$. Assume that $k_d$ is odd. This algorithm outputs a polynomial $f\in \QQx$ such that $f - h^2$ is a sum of $4$ \textup(or fewer\textup) squares.
%\begin{enumerate}
%\item\label{st:reduce_helper:sos4} Check \textup(use \cite[Theorem~17.2]{Rajwade1993}\textup) whether $f$ itself is a sum of $4$ squares, if it is output $h := 0$ and quit.
%\item\label{st:reduce_helper:epsilon} Find a positive number $\varepsilon$ such that 
%\[
%\varepsilon < \inf\bigl\{ f(x)\st x\in \RR \bigr\}.
%\]
%\item Set $l_1 := \bigl\lceil -\sfrac12\cdot \lg \varepsilon\bigr\rceil$.
%\item Set $l_2 := \lceil -\sfrac{k_0}{2}\rceil + 1$.
%\item Set 
%\[
%l_3 := \left\lceil \max \Bigl\{
%\frac{jk_d - dk_j}{2d - 2j}
%\st 0 < j < d
%\Bigr\}\right\rceil.
%\]
%\item Initialize $l := \max \{ l_1, l_2, l_3\}$.
%\item\label{st:reduce_helper:loop} While $\gcd(d, 2l + k_d) \neq 1$ increment $l$ by $1$.
%\item Output $h := 2^{-l}$.
%\end{enumerate}
%\end{alg}

\begin{algorithm}
	\caption{Reduction to a sum of 4 squares: odd valuation case}
	\label{alg:reduce_helper}
	\begin{algorithmic}[1]
		\REQUIRE A positive square-free polynomial $f = c_0 + c_1x + \dotsb + c_dx^d\in \QQx$. The $2$-adic valuations of the coefficients of~$f$ are $k_j := \ord_2 c_j$ for $0\leq j\leq d$. Ensure $k_d$ is odd.
	It is assumed that $f$ is not a sum of 4 squares.
		\ENSURE A polynomial $h\in \QQx$ such that $f - h^2$ is a sum of $4$ \textup(or fewer\textup) squares.
%		\IF{$f$ itself is a sum of~$4$ squares \COMMENT{Use \cite[Theorem~17.2]{Rajwade1993} to check it} \label{st:reduce_helper:sos4}} 
%            \RETURN $g_1 := g_2 := 0$.
%        \ENDIF
        \STATE Find a positive number $\varepsilon$ such that 
\[
\varepsilon < \inf\bigl\{ f(x)\st x\in \RR \bigr\}.
\] \label{st:reduce_helper:epsilon}
        \STATE Set $l_1 := \bigl\lceil -\sfrac12\cdot \lg \varepsilon\bigr\rceil$.
        \STATE  Set $l_2 := \lceil -\sfrac{k_0}{2}\rceil + 1$.
        \STATE Set 
\[
l_3 := \left\lceil \max \Bigl\{
\frac{jk_d - dk_j}{2d - 2j}
\st 0 < j < d
\Bigr\}\right\rceil.
\]
        \STATE Initialize $l := \max \{ l_1, l_2, l_3\}$.
        \WHILE{$\gcd(d, 2l + k_d) \neq 1$  \label{st:reduce_helper:loop}}
        \STATE $l:=l+1$.
        \ENDWHILE
		\RETURN  $h := 2^{-l}$.
	\end{algorithmic}
\end{algorithm}

%\begin{alg}\label{alg:reduce}
%Given a positive square-free polynomial $f = f_0 + f_1x + \dotsb + f_dx^d\in \QQx$, this algorithm outputs polynomials $g_1, g_2\in \QQx$ such that $f - g_1^2 - g_2^2$ is a sum of $4$ \textup(or fewer\textup) squares.
%\begin{enumerate}
%\item Use \cite[Theorem~17.2]{Rajwade1993} to check whether $f$ itself is a sum of~$4$ squares. If it is output $g_1 := g_2 := 0$ and quit.
%\item Denote the $2$-adic valuations of the constant term and the leading coefficient of~$f$ by $k_0 := \ord_2 f_0$ and $k_d := \ord_2 f_d$, respectively.
%\item\label{st:reduce:kd} If $k_d$ is odd, then execute Algorithm~\ref{alg:reduce_helper}. Denote its output by~$h$. Output $g_1 := h$, $g_2 := 0$ and quit.
%\item\label{st:reduce:k0} If $k_0$ is odd then:
%    \begin{enumerate}
%    \item Set $f_* := f_d + f_{d-1}x + \dotsb + f_0x^d$.
%    \item\label{st:reduce:reciprocal} Execute Algorithm~\ref{alg:reduce_helper} for~$f_*$ and denote its output by~$h$.
%    \item\label{st:reduce:reciprocal_output} Output $g_1 := x^{\sfrac{d}{2}} \cdot h(\sfrac1x)$, $g_2 := 0$ and quit.
%    \end{enumerate}
%\item\label{st:reduce:even} Otherwise, if $k_0$ and $k_d$ are both even, then:
%    \begin{enumerate}
%    \item Execute Algorithm~\ref{alg:reduce_helper} for~$2f$ and denote its output by~$h$.
%    \item Output $g_1 := g_2 := \sfrac{h}{2}$ and quit.
%    \end{enumerate}
%\end{enumerate}
%\end{alg}

\begin{algorithm}
	\caption{Reduction to a sum of 4 squares: general case}
	\label{alg:reduce}
	\begin{algorithmic}[1]
		\REQUIRE A positive square-free polynomial $f = c_0 + c_1x + \dotsb + c_dx^d\in \QQx$, that is a priori known not to be a sum of 4 squares.
		\ENSURE Polynomials $g_1,g_2\in \QQx$ such that $f - g_1^2 - g_2^2$ is a sum of $4$ \textup(or fewer\textup) squares.
%		\IF{$f$ itself is a sum of~$4$ squares \COMMENT{Use \cite[Theorem~17.2]{Rajwade1993} to check it}} 
%            \RETURN $g_1 := g_2 := 0$.
%        \ENDIF
        \STATE Denote the $2$-adic valuations of the constant term and the leading coefficient of~$f$ by $k_0 := \ord_2 c_0$ and $k_d := \ord_2 c_d$, respectively.
        \IF{$k_d$ is odd }
            \STATE Execute Algorithm~\ref{alg:reduce_helper}. Denote its output by~$h$.
            \RETURN $g_1 := h$, $g_2 := 0$. \label{st:reduce:kd}
        \ELSIF{$k_0$ is odd}
            \STATE Set $f_* := c_d + c_{d-1}x + \dotsb + c_0 x^d$.\label{st:reduce:reciprocal}
            \STATE Execute Algorithm~\ref{alg:reduce_helper} for~$f_*$ and denote its output by~$h$. 
            \RETURN $g_1 := x^{\sfrac{d}{2}} \cdot h(\sfrac1x)$, $g_2 := 0$. \label{st:reduce:reciprocal_output}
        \ELSE
            \STATE Execute Algorithm~\ref{alg:reduce_helper} for~$2f$ and denote its output by~$h$. \label{st:reduce:even}
            \RETURN $g_1 := \sfrac{h}{2}$, $g_2 := \sfrac{h}{2}$.        
		\ENDIF
	\end{algorithmic}
\end{algorithm}

\begin{prop}
Let $f$ be a positive square-free polynomial in $\QQx$
of degree $d$
with coefficients of heights $\leq H.$
Let us assume that $f$ is not a sum
of $4$ \textup(or fewer\textup) squares.
Then Algorithm \ref{alg:reduce},
relying on Algorithm \ref{alg:reduce_helper},
computes polynomials $g_1,g_2\in \QQx$ such that $f - g_1^2 - g_2^2$ is a sum of $4$ \textup(or fewer\textup) squares.

The binary complexity and the bitsize of the outputs
are polynomials in $d$, $H,$ 
$\lg (\inf f)$ and $\lg (\inf f_*).$ 
\end{prop}

\begin{poc}[Algorithm \ref{alg:reduce_helper}]
We have assumed that $f$ is not a sum of four squares.
%Assume that $f$ is not a sum of four squares so that the algorithm proceeds beyond line~\ref{st:reduce_helper:sos4}. 
Let us first observe that the loop in line~\ref{st:reduce_helper:loop} terminates. Indeed, $k_d$ is odd by assumption so by Dirichlet's prime number theorem the arithmetic progression $\bigl(k_d + 2l\st l\geq \max\{l_1, l_2, l_3\}\bigr)$ contains infinitely many prime numbers. In particular it must contain a number relatively prime to the degree~$d$. 
Moreover, $\bigl( k_d + 2l \mod d \bigr)$ has period $\leq d$
so the Loop in Line \ref{st:reduce_helper:loop} is executed at most $d$
times and the Algorithm terminates. 

Let $h$ be the polynomial constructed by the algorithm. We claim that $f - h^2$ is irreducible in $\QQ_2[x]$. We have $l\geq l_2 > -\sfrac{k_0}{2}$, hence the $2$-adic valuation of the constant term of $f - h^2$ is 
\[
\ord_2\bigl(c_0 - 2^{-2l}\bigr) = \min\{k_0, -2l\} = -2l. 
\]
All the other coefficients of $f - h^2$ coincide with the corresponding coefficients of~$f$. The condition $l\geq l_3$ implies that for every $j \in \{1, \dotsc, d-1\}$ the point $(j, k_j)$ lies on or above the line segment with the endpoints $(0,-2l)$ and $(d, k_d)$. Thus, the Newton polygon of $f - h^2$ consists of just this single line segment, whose slope is $\sfrac{(2l + k_d)}{d}$. Now, $d$ and $2l + k_d$ are relatively prime, hence $f - h^2$ is irreducible in $\QQ_2[x]$ by the well known property of the Newton polygon (see, e.g., \cite[Lemma~3.5]{CG2000}). This proves our claim. 

Finally, the polynomial $f$ being positive and square-free cannot have any real roots. Hence it is separated from zero, which means that one can find $\varepsilon > 0$ as in line~\ref{st:reduce_helper:epsilon}. Now, the condition $l\geq l_1\geq \lceil -\sfrac12\cdot \lg \varepsilon\rceil$ implies that $f - h^2 = f - 2^{-2l}$ is positive, too. Consequently, \cite[Theorem~17.2]{Rajwade1993} asserts that $f - h^2$, being positive and irreducible in $\QQ_2[x]$, is a sum of four (or fewer) squares. 
\end{poc}

\begin{poc}[Algorithm \ref{alg:reduce}]
First, observe that~$f$, being positive, cannot have any real roots. In particular, its constant term $c_0$ must be nonzero. Now, if $k_d$ is odd then the correctness of line~\ref{st:reduce:kd} follows from the already proven correctness of Algorithm~\ref{alg:reduce_helper}. Now, suppose that it is the constant term of~$f$ that has an odd $2$-adic valuation. The reciprocal polynomial $f_*$ constructed in line~\ref{st:reduce:reciprocal} satisfies the identity 
\[
f_*(x) = x^d\cdot f\bigl(\sfrac1x\bigr). 
\]
Thus, it is positive hence a sum of squares, and its leading coefficient is~$c_0$, which has an odd $2$-adic valuation. Consequently, using Algorithm~\ref{alg:reduce_helper} we can find $h\in \QQx$ such that $g := f_* - h^2$ is a sum of $4$ squares. Say $g = a_1^2 + \dotsb + a_4^2$ . Let $g_1$ be given as in line~\ref{st:reduce:reciprocal_output}. We have
\begin{align*}
f - g_1^2
= & \ x^d\cdot f_*\Bigl(\frac1x\Bigr) 
    - \left( x^{\frac{d}{2}}\cdot h\Bigl(\frac1x\Bigr)\right)^2
=  x^d\cdot g\Bigl(\frac1x\Bigr) \\
= & \ \left( x^{\frac{d}{2}}\cdot a_1\Bigl(\frac1x\Bigr)\right)^2 
    + \dotsb + \left( x^{\frac{d}{2}}\cdot a_4\Bigl(\frac1x\Bigr)\right)^2
\end{align*}
is a sum of four squares, as well. This proves the correctness of line~\ref{st:reduce:reciprocal_output}. The correctness of the last case, when both $k_0$ and $k_d$ are even, is trivial.
\end{poc}
\begin{comp_analysis}[Algorithm \ref{alg:reduce_helper}][Algorithm \ref{alg:reduce}]
For Algorithm \ref{alg:reduce_helper},
the desired $\varepsilon$ can be computed
by checking whether $f-2^{2^s}$
has any real roots, for increasing $s.$
This can be done using classical methods like
Descartes' rule of sign or the computation
of signatures of special Hankel matrices defined
by the coefficients of $f-2^{2^s}$.
This is polynomial in $d$, the heights of the $c_j$'s 
and $\lg (\inf f)$.
The same is true for the remaining operations.
The output $h$ has size polynomial in these quantities.
Then, it is clear that 
the complexity of Algorithm \ref{alg:reduce} is completely given by that of Algorithm \ref{alg:reduce_helper}.
\end{comp_analysis}

\begin{rem}
Algorithm~\ref{alg:reduce_helper} always outputs a constant polynomial, hence in line~\ref{st:reduce:reciprocal_output} above we actually have $g_1 = x^{\sfrac{d}{2}}\cdot h(\sfrac1x) = x^{\sfrac{d}{2}}\cdot h$. We use the ``baroque'' notation $h(\sfrac1x)$ to allow one to substitute Algorithm~\ref{alg:reduce_helper} with any other algorithm having the same input and output specification.
\end{rem}

\section{Final algorithm: sums of six squares}
\label{sec:sos6}

We are now ready to present our main algorithm that decomposes any nonnegative univariate polynomial in $\QQx$ into a sum of six squares of polynomials in $\QQx$.
Its correctness is clear thanks to the correctness of the previous algorithms.
%\begin{alg}
%\label{alg:sixsquares}
%Given a nonzero nonnegative polynomial $f\in \QQx$, this algorithm outputs polynomials $f_1, \dotsc, f_6\in \QQx$ such that $f_1^2 + \dotsb + f_6^2 = f$.
%\begin{enumerate}
%\item If $f$ is a square, then output $f_1 := \sqrt{f}$, $f_2 := \dotsb f_6 := 0$ and quit.
%\item Check if $f$ is a sum of~$2$ squares \textup(see Observation~\ref{obs:sos2}\textup). If it is, then: 
%    \begin{enumerate}
%    \item Execute Algorithm~\ref{alg:sos2} to obtain polynomials $f_1$, $f_2$ such that $f_1^2 + f_2^2 = f$.
%    \item Output $f_1, f_2$ and $f_3 := \dotsb f_6 := 0$ and quit.
%    \end{enumerate}
%\item Check if $f$ is a sum of~$4$ squares \textup(see \cite[Theorem~17.2]{Rajwade1993}). If it is, then: 
%    \begin{enumerate}
%    \item Execute Algorithm~\ref{alg:sos4}, to obtain polynomials $f_1, \dotsc, f_4$ such that $f_1^2 + \dotsb + f_4^2 = f$.
%    \item Output $f_1, \dotsc, f_4$ and $f_5 := f_6 := 0$ and quit.
%    \end{enumerate}
%\item Using square-free decomposition, factor~$f$ into a product $gh^2$, where $g$ is square-free.
%\item Execute Algorithm~\ref{alg:reduce} with~$g$ as an input to obtain polynomials $g_1, g_2\in \QQx$ such that $g - g_1^2 - g_2^2$ is a sum of~$4$ squares.
%\item Execute Algorithm~\ref{alg:sos4} to decompose $g - g_1^2 - g_2^2$ into a sum of~$4$ squares. Denote the output by $g_3, \dotsc, g_6$.
%\item Output $f_1 := g_1h, \dotsc, f_6 := g_6h$
%\end{enumerate}
%\end{alg}

\begin{algorithm}
	\caption{Decomposition of a nonnegative univariate rational polynomial into a sum of 6 squares}
	\label{alg:sixsquares}
	\begin{algorithmic}[1]
		\REQUIRE A nonnegative polynomial $f \in  \QQx$. 
		\ENSURE Polynomials $f_1, \dotsc, f_6\in \QQx$ such that $f_1^2 + \dotsb + f_6^2 = f$.
		\IF{$f$ is a square}
            \RETURN $f_1 := \sqrt{f}$, $f_2 := \dotsb f_6 := 0$.
        \ENDIF
		\IF{$f$ is a sum of~$2$ squares \COMMENT{Use Observation~\ref{obs:sos2} to check it}}
            \STATE Execute Algorithm~\ref{alg:sos2} to obtain $f_1$, $f_2 \in \QQx$ such that $f_1^2 + f_2^2 = f$.
            \RETURN $f_1, f_2$ and $f_3 := \dotsb f_6 := 0$.
        \ENDIF
        \IF{$f$ is a sum of~$4$ squares \COMMENT{Use \cite[Theorem~17.2]{Rajwade1993} to check it}} 
            \STATE Execute Algorithm~\ref{alg:sos4}, to obtain $f_1, \dotsc, f_4 \in \QQx$ such that $f_1^2 + \dotsb + f_4^2 = f$.
            \RETURN $f_1, \dotsc, f_4$ and $f_5 := f_6 := 0$
        \ENDIF        
		\STATE Compute the square-free decomposition of $f = g\cdot h^2$, where $g,h \in \QQx$ and $g$ is square-free.
        \STATE Execute Algorithm~\ref{alg:reduce} with~$g$ as an input to obtain $g_1, g_2\in \QQx$ such that $g - g_1^2 - g_2^2$ is a sum of~$4$ squares in $\QQx$.
        \STATE Execute Algorithm~\ref{alg:sos4} to decompose $g - g_1^2 - g_2^2$ into a sum of~$4$ squares in $\QQx$. Denote the output by $g_3, \dotsc, g_6$.
        \RETURN $f_1 := g_1h, \dotsc, f_6 := g_6h$.
	\end{algorithmic}
\end{algorithm}

\begin{thm}
If $f \in \QQx$ is a nonnegative univariate polynomial
of degree $d$ with coefficients of heights $\leq H$,
then Algorithm \ref{alg:sixsquares}
computes a decomposition of $f$ as a sum of 
at most $6$ squares. \\
There exists some 
$H_1=\poly (d,H,\inf (f), \inf (f_*))$
such that the expected binary complexity is 
polynomial in
$\IntFact(H)$, $\PolyFact(d,H_1)$, 
$\UGComp(2d,H_1)$, and $\SFUnits(2d,H_1)$. 
The output size is polynomial in $\SFUnits(2d,H_1).$ 
\end{thm}
%\begin{poc}[Algorithm \ref{alg:sixsquares}]
%Clear thanks to the correctness of the 
%previous algorithms.
%\end{poc}
\begin{comp_analysis}[Algorithm \ref{alg:sixsquares}]
First of all,
we should remark that checking
whether whether $f$ is a sum
of 2 or 4 squares (by Observation \ref{obs:sos2} or
\cite[Theorem~17.2]{Rajwade1993} respectively) is
at worse as costly as computing such a decomposition.
Hence the total complexity will be given by
the cost of a decomposition into a sum of 2
squares, a sum of 4 squares and a polynomial factorization.

If $f$ is of degree $d$ with coefficients of
heights $\leq H$, then there is some
$H_1=\poly (d,H,\inf (f), \inf (f_*))$, given by Algorithm \ref{alg:sos2},  Algorithm \ref{alg:reduce} and the $u \mapsto 2ud+8d(1+\lg (d))$ mapping
such that the total expected binary cost
is polynomial in
$\IntFact(H)$, $\PolyFact(d,H_1)$,  $\UGComp(2d,H_1)$ and $\SFUnits(2d,H_1)$. 
The output size is polynomial in $\SFUnits(2d,H_1).$ 
\end{comp_analysis}

%\section{Bit complexity analysis}
%\label{sec:complexity}

\begin{rem}
In Algorithm \ref{alg:reduce}, either $g_2=0$
or $g_1=g_2$.
It means that for some $a_5=1$ or $2,$ the output of Algorithm \ref{alg:sixsquares}
is of the form
\[ f=f_1^2+f_2^2+f_3^2+f_4^2+a_5f_5^2.\]
In other words, the output is a \textit{weighted sum of 5 squares}.
\end{rem}

\begin{rem}
It is possible to adapt Algorithms \ref{alg:sos2}
and \ref{alg:sos4} to output 
\textit{weighted sum of squares}
of the form $f=\lc(f)f_1^2+\lc(f)f_2^2$ or $f=\lc(f)f_1^2+\dots+\lc(f)f_4^2$.
Using the previous remark, one can then adapt  
Algorithm \ref{alg:sixsquares}
to output a \textit{weighted sum of 5 squares}
\[f=a_1 f_1^2+\dots+a_5 f_5^2\] with the $a_i$'s being nonnegative
integers.
One major benefit if we do so is that the adapted Algorithm \ref{alg:sos2}
will then need no
integer factorization of $\lc(f),$
and we then need no reference to $\IntFact.$
\end{rem}

%\section*{Appendix}

Algorithm \ref{alg:sixsquares} is sub-optimal in the sense that it produces \textbf{six} squares, while it is known that only \textbf{five} are really needed. 
Below, we present Algorithm \ref{alg:conj_reducing_to_4_squares}, which is a variant of Algorithm~\ref{alg:reduce_helper}. 

%\begin{alg}
%Given a nonzero, square-free positive polynomial $f = f_0 + f_1x + \dotsb + f_dx^d\in \QQx$, this algorithm outputs a polynomial $h\in \QQx$ such that $f - h^2$ is a sum of~$4$ \textup(or fewer\textup) squares.
%\begin{enumerate}
%\item If $f$ itself is a sum of $4$ squares, then output $h := 0$ and quit.
%\item Denote $f_* := f_d + f_{d-1}x + \dotsb + f_0x^d$.
%\item Find a positive number $\varepsilon$ such that 
%\[
%\varepsilon < \inf\bigl\{ f(x) \st x\in \RR\bigr\}
%\qquad\text{and}\qquad
%\varepsilon < \inf\bigl\{ f_*(x)\st x\in \RR\bigr\}.
%\]
%\item Initialize $l := \lceil -\sfrac12\cdot \lg\varepsilon\rceil$.
%\item Repeat the following steps: 
%    \begin{enumerate}
%    \item If $f - 2^{-2l}$ is irreducible in $\QQ_2[x]$, then output $h := 2^{-l}$ and quit. 
%    \item If $f - 2^{-2l} x^d$ is irreducible in $\QQ_2[x]$, then output $h := 2^{-l}x^{\sfrac{d}{2}}$ and quit. 
%    \item Otherwise increment $l$ by~$1$ and reiterate the loop.
%    \end{enumerate}
%\end{enumerate}
%\end{alg}

\begin{algorithm}
	\caption{Reduction to a sum of 4 squares}
	\label{alg:conj_reducing_to_4_squares}
	\begin{algorithmic}[1]
		\REQUIRE A positive square-free polynomial $f = c_0 + c_1x + \dotsb + c_dx^d\in \QQx$. 
		\ENSURE A polynomial $h\in \QQx$ such that $f - h^2$ is a sum of~$4$ \textup(or fewer\textup) squares.
		\IF{$f$ is a sum of $4$ squares}
            \RETURN $h := 0$.
        \ENDIF       
		\STATE Set $f_* := c_d + c_{d-1}x + \dotsb + c_0x^d$.
        \STATE Find a positive number $\varepsilon$ such that 
\[
\varepsilon < \inf\bigl\{ f(x) \st x\in \RR\bigr\}
\quad\text{and}\quad
\varepsilon < \inf\bigl\{ f_*(x)\st x\in \RR\bigr\}.
\]
        \STATE Initialize $l := \lceil -\sfrac12\cdot \lg \varepsilon\rceil$.
        \WHILE{True}
            \IF{$f - 2^{-2l}$ is irreducible in $\QQ_2[x]$}
                \RETURN $h := 2^{-l}$.
            \ENDIF
            \IF{$f - 2^{-2l}x^d$ is irreducible in $\QQ_2[x]$}
                \RETURN $h := 2^{-l}x^{\sfrac{d}{2}}$.  
            \ENDIF
            \STATE $l:=l+1.$
        \ENDWHILE        
	\end{algorithmic}
\end{algorithm}

We verified Algorithm \ref{alg:conj_reducing_to_4_squares} empirically on over $20 \, 000$ random nonnegative univariate polynomials. For all of them, it worked fine. Yet still, we don't know how to prove that it eventually terminates, except when either $\ord_2 c_0$ or $\ord_2 c_d$ is odd, in which cases it reduces to Algorithm~\ref{alg:reduce_helper}. 
On the other hand, the correctness of the output of the algorithm is immediate. Hence, if we can prove that it stops, we can use it to get the desired decomposition of any nonnegative polynomial into a sum of five (instead of six) squares.
This is the first further research direction that is left open by this work. 
We also intend to compare our algorithms with the other ones analyzed in \cite{magron2019algorithms}.

%\bibliographystyle{ACM-Reference-Format}
%\bibliography{sos_biblio}

\end{document}